\documentclass[12pt, draftclsnofoot, onecolumn]{IEEEtran}
\usepackage{graphicx}
\usepackage{epstopdf}
\usepackage{cite}
\usepackage{amsfonts}
\usepackage{amssymb}
\usepackage{amsmath}
\usepackage{mathrsfs}
\usepackage{bm}
\usepackage{amsmath}
\usepackage{subfigure}
\usepackage{mdwmath}
\usepackage{mdwtab}
\usepackage{enumerate}
\usepackage{dsfont}

\DeclareMathOperator*{\argmin}{arg\,min}

\newtheorem{theorem}{Theorem}

\newtheorem{lemma}{Lemma}

\begin{document}

	\title{Optimal Sleeping Mechanism for Multiple Servers with MMPP-Based Bursty Traffic Arrival}
	\author{Zhiyuan Jiang, Bhaskar Krishnamachari, Sheng Zhou, \\Zhisheng Niu,~\IEEEmembership{Fellow,~IEEE}
	\thanks{
    Z. Jiang, S. Zhou and Z. Niu are with Tsinghua National Laboratory for Information Science and Technology, Tsinghua University, Beijing 100084, China. Emails: \{zhiyuan, sheng.zhou, niuzhs\}@tsinghua.edu.cn. 
    
    B. Krishnamachari is with the Ming Hsieh Department of Electrical Engineering, University of Southern California, Los Angeles, CA 90089, USA. Email: bkrishna@usc.edu.
  
    This work is sponsored in part by the Nature Science Foundation of China (No. 61701275, No. 91638204, No. 61571265, No. 61621091), the China Postdoctoral Science Foundation, and Hitachi R\&D Headquarter.
    }
    }
	\maketitle
	
	\vspace{-10mm}
	\begin{abstract}
    An important fundamental problem in green communications and networking is the operation of servers (routers or base stations) with sleeping mechanism to optimize energy-delay tradeoffs. This problem is very challenging when considering realistic bursty, non-Poisson traffic. We prove for the first time the optimal structure of such a sleep mechanism for multiple servers when the arrival of jobs is modeled by a bursty Markov-modulated Poisson process (MMPP). It is shown that the optimal operation, which determines the number of active (or sleeping) servers dynamically, is hysteretic and monotone, and hence it is a queue-threshold-based policy. This work settles a conjecture in the literature that the optimal sleeping mechanism for a single server with interrupted Poisson arrival process, which can be treated as a special case of MMPP, is queue-threshold-based. The exact thresholds are given by numerically solving the Markov decision process.    
	\end{abstract}
	
	\section{Introduction}
	\label{sec_intro}
	The reduction of \emph{energy consumption} has attracted more and more attention in several engineering fields, e.g., wireless communication systems and data centers. One of the most effective approaches is to put idle servers into sleeping mode due to the fact that a significant amount of energy is wasted by keeping the idle servers active. Concretely, a base station (BS) consumes $90\%$ of its peak power even when the traffic load is low \cite{oh11}, and a typical idle server consumes $50\%$-$60\%$ of its peak power. On the other hand, the utilization of BSs and servers is usually low, especially with more and more densely deployed infrastructures \cite{kim10}.  	
	Meanwhile, the energy consumption reduction thanks to sleeping operations comes with an undesired user \emph{delay} increase, due to the extra job queuing time with possibly sleeping servers. Therefore, the design of sleeping mechanism should consider the tradeoff between energy consumption and queuing delay, and in the meantime avoid frequent server mode switching which costs extra energy.
	
	In practice, the arrival traffic at servers often exhibits a high level of \emph{burstness} \cite{niu11}, whereas existing works usually focus on Poisson-based, non-bursty traffic arrivals. It is important to understand the impact of traffic burstness since, intuitively, it may create more sleeping opportunities. However, the optimization of the energy-delay tradeoff with bursty, non-Poisson traffic becomes very challenging and hence few results are available.
	
	\subsection{Related Work and Main Contributions}
	The BS and server sleeping mechanisms have attracted wide attention in the literature. It is proved by Kamitsos \emph{et al.} \cite{kam10} that the optimal structure of the sleeping operations for Poisson arrival and a single server is queue-threshold-based. The proof is built upon the previous work by Lu, Serfozo \cite{lu84} and Hipp, Holzbaur \cite{sab88}. In the work by Wu \emph{et al.} \cite{wu16} and Leng \emph{et al.} \cite{bing17}, the arrival traffic pattern is generalized to interrupted Poisson process (IPP) to capture the traffic burstness. In an IPP process, jobs only arrive during the ON phase and the ON and OFF phases transit to each other based on a Markov process. Specifically, Wu \emph{et al.} \cite{wu16} calculate the optimum queue threshold with IPP arrival and a single server by fixing the sleeping policy to be $N$-based, i.e., turning on the server when there are $N$ jobs and turning it off when the queue is empty. However, it is shown in the work by Leng \emph{et al.} \cite{bing17} that the optimal sleeping policy with IPP arrival has two sets of thresholds, meaning that in each phase of the IPP the thresholds to turn on and off the server are different. Therefore, the $N$-based policy \cite{wu16} is in general not optimal with IPP arrival. Towards finding the optimal sleeping policy, Leng \emph{et al.} \cite{bing17} adopt a partially observable Markov decision process (POMDP) formulation and analyze the optimal sleeping policy with IPP arrivals and a single server numerically. It is proved that the optimal policy is hysteretic but the monotonicity property, which together with the hysteretic property proves the optimal policy structure to be queue-threshold-based, is left as a conjecture. 
	
	In this letter, we generalize the existing work by considering the Markov-modulated-Poisson-process (MMPP) traffic arrival and multiple servers. The optimal sleeping policy structure is proved to be queue-threshold-based, and hence the conjecture by Leng \emph{et al.} \cite[Conjecture 1]{bing17} is settled since IPP and a single server can be considered as a special case. Numerical results are also given to shed light upon the optimum queue thresholds.
	
	\section{System Model and Problem Formulation}
	\label{sec_model}
	
	We consider $M$ servers, each has two operation modes, active and sleeping. The $M$ servers serve jobs in a single queue with a buffer size of $B$. Denote the number of active servers as $W$, and $W \in \{0,...,M\}$. We assume that jobs arrive at the queue according to an MMPP to capture the burstiness of the traffic. Jobs arrive during the $S$-th phase of MMPP based on the Poisson process with rate $\lambda_S$, where the MMPP arrival phase is denoted as $k_S$ and $S \in \{1,...,N\}$. The MMPP is parameterized by the $N$-state continuous time Markov chain with phase transition matrix as $\bm{R} = \left[ {\begin{array}{*{20}{c}}
    { - {\sigma _1}}& \cdots &{{\sigma _{1N}}}\\
     \vdots & \ddots & \vdots \\
    {{\sigma _{N1}}}& \cdots &{ - {\sigma _N}}
    \end{array}} \right]$,
	where $\sigma_{S_1 S_2}$ denotes the transition rate from phase $k_{S_1}$ to $k_{S_2}$ of MMPP, and $\sigma_{i}=\sum_{j=1,\,j \neq i}^N{\sigma_{ij}}$. The service time is assumed to be independently and identically distributed according to an exponential distribution over jobs with mean service time of $\mu^{-1}$ for each active server. Based on the queuing theory, the service rate for $W$ active servers is $W \mu$. The memory-less property of the arrival\footnote{Although the arrival MMPP is not a renewal process, the arrival phase transition is still memory-less based on the MMPP definition.} and departure processes enables us to formulate the problem as a continuous-time MDP. The system state is denoted as $(S,Q,W)$, where $S \in \{1,...,N\}$, $W \in \{0,...,M\}$ and $Q\in\{0,\cdots, B\}$. The state 
	$(S,Q,W)$ denotes that there are $Q$ jobs in the queue, the number of active servers is $W$ and the arrival MMPP is in the $S$-th phase. The control action space is $\{0,...,M\}$, wherein an action $u_a$ turns $a$ servers to the active mode.\footnote{Obviously, considering the switching cost, it is better to turn an additional $a-W$ servers to active mode when there are $W$ ($W \le a$) active servers, rather than to close some servers and turn on more.}  In the case of $a$ is smaller than the number of current active servers ($W$), the action means to turn $W-a$ servers to the sleeping mode.
	
	We adopt the discrete-time approximation of the continuous-time MDP, whereby the time is divided into time slots and time duration of each time slot, i.e., denoted by $\Delta$, is sufficiently small such that there is at most one event (job arrival, departure, or arrival phase shift) occurrence in one time slot \cite[Chapter 5.5]{gal12}. The decision is made at each time slot, and the time index is conveyed in the brackets. The system states evolve as follows
	\begin{IEEEeqnarray}{rCl}
		\label{eqn:wsq}
		&W(t+1)&=a(t),\\
		&S(t+1)&=\left\{ {\begin{array}{ll}
				\bar{S}, & \textrm{if arrival phase transits to $k_{\bar{S}}$ phase}; \\
				S(t),   & \textrm{no phase transition happens},
				\end{array}} \right.\\
		&Q(t+1)&=\left\{ {\begin{array}{ll}
				Q(t)+1, & \textrm{if } Q(t)<B \textrm{ and a job arrives}; \\
				Q(t)-1, & \textrm{if } Q(t)>0 \textrm{ and a job is served} ;\\
				Q(t),   & \textrm{otherwise}.
				\end{array}} \right. 
	\end{IEEEeqnarray}	
	The state transition probability given action $u_a$ is (the time index is omitted for simplicity)
	\begin{equation}
		\label{prob}
		\textrm{Pr}\{(S,Q,W)\to(S,Q+1,a)\}=\lambda_S \Delta \mathds{1}(Q < B). 
	\end{equation}	
	\begin{equation}
		\textrm{Pr}\{(S,Q,W)\to(S,Q-1,a)\}=a \mu \Delta \mathds{1}(Q > 0). 
	\end{equation}	
	\begin{equation}
		\textrm{Pr}\{(S,Q,W)\to(\bar{S},Q,a)\}=\sigma_{S \bar{S}} \Delta,\, \bar{S} \in \mathcal{N}_S .
	\end{equation}	
	\begin{IEEEeqnarray}{rCl}
		\textrm{Pr}\{(S,Q,W)\to(S,Q,a)\} = 1-a \mu \Delta \mathds{1}(Q > 0) - \lambda_S \Delta \mathds{1}(Q < B) -\sum_{\bar{S} \in \mathcal{N}_S }\sigma_{S \bar{S}}\Delta,
	\end{IEEEeqnarray}
	where $\mathcal{N}_S = \{1,...,N\} \backslash \{S\}$. All other transition probabilities are zeros.
	
	We consider the active energy consumption cost, switching energy cost, and delay cost of the system. The objective is to minimize the total discounted cost \cite{lu84,wu16,bing17}, i.e., 
	\begin{IEEEeqnarray}{rCl}
		\label{minE}
        \min_{a(t) \in \{0,...,M\}}\mathbb{E}\left[\sum_{t=1}^{\infty} \right. && r^{t-1}(\max(a(t)-W(t),0)E_{\textrm{sw}}
         + \omega Q(t) + a(t)E_{\textrm{on}}\Bigg],
	\end{IEEEeqnarray}	
	where $r \in [0,1)$ is the discount factor which reflects how important is immediate cost, the tradeoff between delay and energy cost is represented by $\omega$, the switching energy consumption is denoted as $E_\textrm{sw}$, and $E_\textrm{on}$ denotes the energy consumption of the server being active for one time slot. Although only the server start-up energy consumption is considered for switching energy cost based on real systems, the inclusion of shut-down cost would not affect the results since a start-up is always followed by a shut-down to complete a busy cycle.

	\section{Optimal Policy Structure}
	\label{sec_opt}	
	
	In Theorem 1 of the work by Leng \emph{et al.} \cite{bing17}, it is proved that the optimal policy with IPP arrival is a hysteretic policy, i.e., if the policy chooses to switch to a better mode, then it would stay in that mode if it is already in the mode. The extension of the hysteretic property to the MMPP case is actually straightforward given the work by Hipp and Holzbaur \cite[Theorem 1]{sab88} and examining the switching cost function which is defined as
	\begin{equation}
	    s(W(t),a(t)) = \max(a(t)-W(t),0) E_\textrm{sw}.
	\end{equation}
	
	It is conjectured by Leng \emph{et al.} \cite[Conjecture 1]{bing17} that the optimal policy for IPP arrival is also a monotone policy, i.e., given $S(t)$ the optimal action $a^{\ast}(t) \triangleq f(S(t),Q(t),W(t))$ is non-decreasing with $Q(t)$. Consequently, assuming the conjecture is upheld, it is shown that the optimal policy for IPP arrival is a threshold-based policy, which is described by the \emph{active} and \emph{sleeping} thresholds at ON and OFF phases of IPP, respectively. In what follows, we not only settle the monotone conjecture, but also extend to MMPP arrival case, and thus prove the optimal policy with MMPP arrival and multiple servers is queue-threshold-based.
	\begin{theorem}
	\label{thm1}
		The optimal policy to the formulated MDP is a monotone policy, i.e., $\forall$ $S$, $W$, and $Q_1\geq Q_2$, 
		\begin{equation}
		f(S,Q_1,W) \geq f(S,Q_2,W).
		\end{equation}
	\end{theorem}
	\begin{IEEEproof}
	The main technique to prove the theorem is inspired by the proof of Theorem 1 in the work by Lu and Serfozo \cite{lu84}. However, the arrival process is Poisson-based and the cost-to-go function is required to be submodular \cite{lu84}. In fact, it can be shown through numerical simulations that the cost-to-go function with MMPP arrival is, in general, not a submodular function. To address this issue, we present a stronger result in Lemma \ref{lm_1} (Appendix) which indicates that only a partial submodular condition is sufficient, i.e., it suffices that the cost-to-go function is submodular with respect to $Q$ and $a$. Define the cost-to-go function as 
    \begin{IEEEeqnarray}{rCl}
	\label{Vt}
        V_t(S,Q,W) &=& \min_{a\in\{0,...,M\}}\{s(W,a)+w_{t-1}(S,Q,a)\}, \\
	    w_t(S,Q,a) &=& \omega Q+a E_\textrm{on} + r\left[\sum_{\bar{S} \in \mathcal{N}_S }\sigma_{S \bar{S}}\Delta V_t(\bar{S},Q,a)\right.\nonumber\\
	    && +\lambda_S \Delta \mathds{1}(Q<B)V_t(S,Q+1,a) +a \mu \Delta \mathds{1}(Q>0)V_t(S,Q-1,a) \nonumber\\
	    && + \Bigg(1-a \mu \Delta \mathds{1}(Q>0) - \lambda_S \Delta \mathds{1}(Q<B)  \left.\left.-\sum_{\bar{S} \in \mathcal{N}_S }\sigma_{S \bar{S}}\Delta\right) V_t(S,Q,a)\right],
	\end{IEEEeqnarray}
    and define 
    \begin{equation}
        u_{t}(S,Q,W,a) \triangleq s(W,a)+w_{t}(S,Q,a).
    \end{equation} 
    To prove Theorem \ref{thm1}, we will first show that Theorem \ref{thm1} is true in a finite horizon of length $T$ by induction. The generalization to infinite horizon follows standard methods as shown by Lu and Serfozo \cite[Theorem 2]{lu84}. In particular, we will show that the following statements are valid.
    
    \begin{enumerate}[(i)]
    \item 
    \label{it1}
    {The optimal policy is non-decreasing in $Q$.}
    \item 
    \label{it2}
    {$\forall S, t$, and $Q_1\leq Q_2$, $W_1\leq W_2$, $V_t(S,Q_2,W_1) - V_t(S,Q_1,W_1) \geq V_t(S,Q_2,W_2)-V_t(S,Q_1,W_2)$.
    }
    \item 
    \label{it3}
    {Define $V'_t(S,Q,W)=V_t(S,Q+1,W)-V_t(S,Q,W)$.\footnote{Increment of other functions over $Q$ is denoted identically.} Then $\forall t, S$ and $Q$, $0\leq V'_t(S,Q,0)\leq V'_t(S,Q+1,M)$.}
    \end{enumerate}
	\textbf{Induction basis:} For $t=1$, the one-step cost-to-go function is
	\begin{IEEEeqnarray}{rCl}
	V_1(S,Q,W)&=&\min_{a_1\in\{0,...,M\}}[\max(a_1-W,0)E_\textrm{sw}  + \omega Q + a_1 E_\textrm{on}].
	\end{IEEEeqnarray}
	It is obvious that the optimal control action $a_1^{\ast}$ to minimize $V_1$ does not depend on $Q$. Therefore, (\ref{it1})-(\ref{it3}) are satisfied with equality.\\
	\textbf{Induction steps:} Suppose (\ref{it1})-(\ref{it3}) are valid for $k\leq t$. Then, for $\forall a\leq b$, and $S$,
   	\begin{IEEEeqnarray}{rCl}
   	\label{eqn:www}
    && w'_t(S,Q,b)-w'_t(S,Q,a)\nonumber\\
    &=& r\left[ \sum_{\bar{S} \in \mathcal{N}_S }\sigma_{S \bar{S}}\Delta (V'_t(\bar{S},Q,b)-V'_t(\bar{S},Q,a))\right. + \lambda_S \Delta (V'_t(S,Q+1,b)-V'_t(S,Q+1,a)) \nonumber\\
    &&+ a \mu \Delta(V'_t(S,Q-1,b)-V'_t(S,Q-1,a)) + \mu (b-a) \Delta(V'_t(S,Q-1,b)-V'_t(S,Q,b)) \nonumber\\
    &&+ \left(1-a \mu \Delta - \lambda_S \Delta -\sum_{\bar{S} \in \mathcal{N}_S }\sigma_{S \bar{S}}\Delta\right)  (V'_t(S,Q,b)-V'_t(S,Q,a))\Bigg] \nonumber\\
   	&\leq& -r\mu (b-a) \Delta(V'_t(S,Q,b)-V'_t(S,Q-1,b)),
   	\end{IEEEeqnarray}
	where the inequality is based on the induction hypothesis (\ref{it2}). Combining the induction hypotheses (\ref{it2}) and (\ref{it3}), it follows that 
	\begin{IEEEeqnarray}{rCl}
	    V'_t(S,Q-1,b) \leq V'_t(S,Q-1,0) &\leq& V'_t(S,Q,M) \leq V'_t(S,Q,b).
	\end{IEEEeqnarray}
	Therefore, we obtain $w'_t(S,Q,b)-w'_t(S,Q,a) \leq 0$, and it follows that $u_t(S,Q,W,a)$ satisfies the conditions in Lemma \ref{lm_1}. Hence, (\ref{it1}) and (\ref{it2}) are proved by noticing that the minimization operation preserves partial submodularity. 
	
	To prove (\ref{it3}), we obtain 
   	\begin{IEEEeqnarray}{rCl}
   	\label{eqn:www2}
    && w'_t(S,Q,0)-w'_t(S,Q+1,M)\nonumber\\
    &=& r \left[\sum_{\bar{S} \in \mathcal{N}_S }\sigma_{S \bar{S}}\Delta (V'_t(S,Q,0)-V'_t(S,Q+1,M))\right. \nonumber\\
    &&+ \lambda_S \Delta (V'_t(S,Q+1,0)-V'_t(S,Q+2,M)) + \mu M \Delta(V'_t(S,Q+1,M)-V'_t(S,Q,M)) \nonumber\\
    &&+ \left(1-a \mu \Delta - \lambda_S \Delta -\sum_{\bar{S} \in \mathcal{N}_S }\sigma_{S \bar{S}}\Delta\right)(V'_t(S,Q,0)   -V'_t(S,Q+1,M))\Bigg] \nonumber\\
   	&\leq& -r \mu M \Delta(V'_t(S,Q,M)-V'_t(S,Q-1,M)) \leq 0,
   	\end{IEEEeqnarray}
	where the inequality stems from combining the induction hypothesis (\ref{it2}) and (\ref{it3}). It follows that $V'_{t+1}$ is non-negative since $V'_t$ is. Denote 
	\begin{equation}
	    x \triangleq f(S,Q,W),\,y \triangleq f(S,Q+2,W),
	\end{equation} 
	we obtain
   	\begin{IEEEeqnarray}{rCl}
   	\label{eqn:vvv2}
    && V'_{t+1}(S,Q,0)\nonumber\\
    &=& \min_{a}u_{t+1}(S,Q+1,W,a)-\min_{a}u_{t+1}(S,Q,W,a) \nonumber\\
    &\leq& u_{t+1}(S,Q+1,W,x)-u_{t+1}(S,Q+1,W,x)\nonumber\\
    &=& w'_t(S,Q,x) \leq w'_t(S,Q,0)\leq w'_t(S,Q+1,M) \leq w'_t(S,Q+1,y) = u^\prime_{t+1}(S,Q+1,M,y)\nonumber\\
    &\leq& V'_{t+1}(S,Q+1,M).
   	\end{IEEEeqnarray}	
	Therefore, the hypothesis (\ref{it3}) is proved. Note that the corner cases wherein $Q=0$ or $Q=B$ can be dealt with appropriately, and for brevity the details are not shown. With this, the induction proof is completed.
    \end{IEEEproof}
    
    \section{Numerical Results}
    \begin{figure}[!t]
    \centering
    \includegraphics[width=0.55\textwidth]{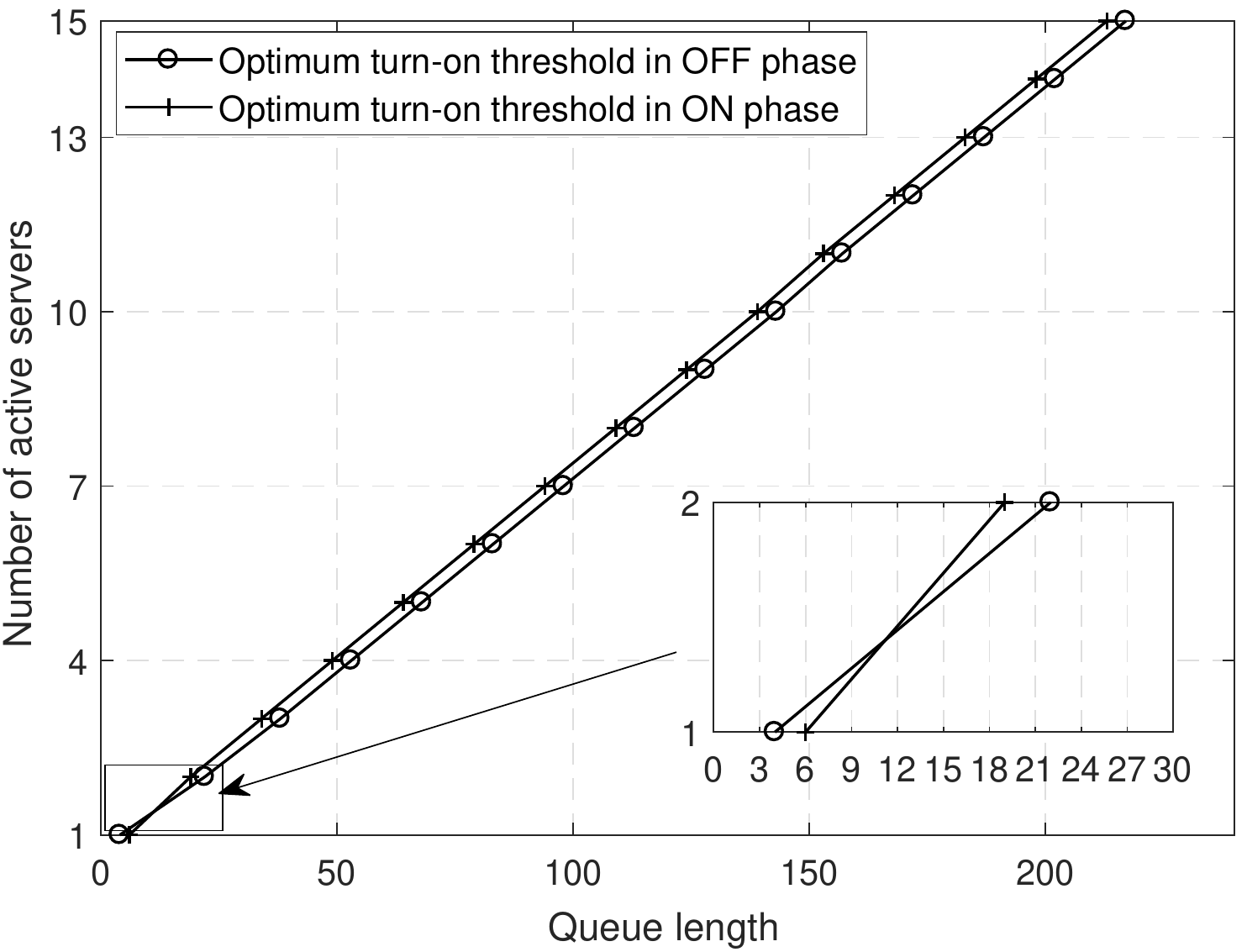}
    \caption{Optimum queue thresholds for turning on servers. }
    \label{fig_on}
    \end{figure}
    \begin{figure}[!t]
    \centering
    \includegraphics[width=0.55\textwidth]{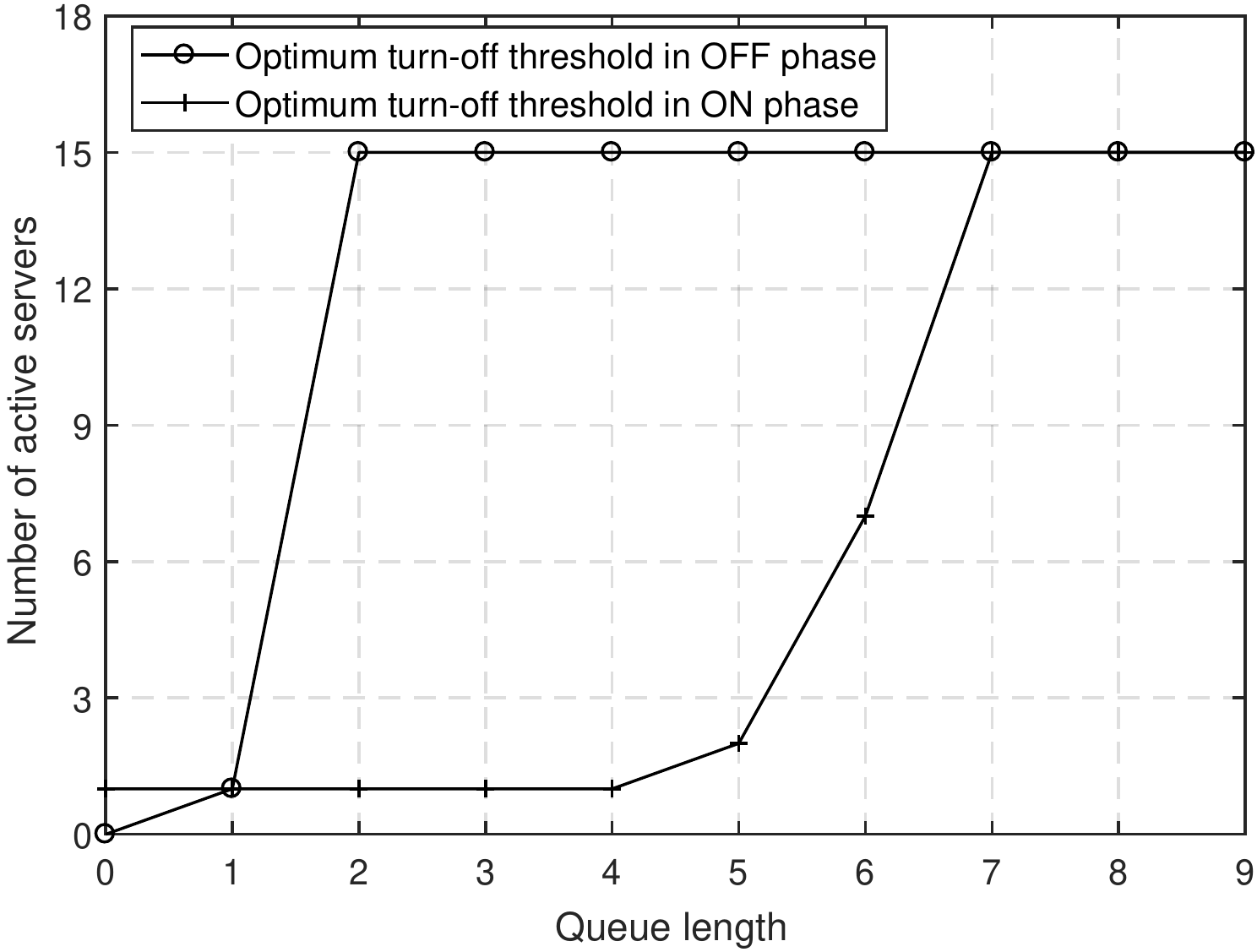}
    \caption{Optimum queue thresholds for turning off servers.}
    \label{fig_off}
    \end{figure}
    In this section, the MDP is solved numerically to obtain the optimum queue thresholds. Each time slot is $10$ milliseconds. Two arrival phases are considered in the MMPP, where the arrival rates are $5$ (ON phase) and $0$ (OFF phase) jobs per second, respectively. The definition of ON and OFF phases is identical with that in the IPP; jobs only arrive during the ON phase based on the Poisson model; the holding time in both phases obeys i.i.d. exponential distributions. The phase transition rates are $0.5$~$s^{-1}$ and $0.25$~$s^{-1}$ in ON phase and OFF phase, respectively. The mean service time for a single server is $0.12$~$s$. The number of available servers is $15$. The buffer size is $250$. The turn-on energy consumption of a server is $200$~joules. The energy consumption of an active server in a time slot is $2.5$ joules. The tradeoff parameter $\omega=0.2$. These parameters are obtained from realistic cellular systems \cite{niu15,bing17}. The discount factor $r=0.999$. The solution to the MDP is obtained by standard policy iterations over infinite horizon. In Fig. \ref{fig_on}, it is shown that the optimum turn-on queue thresholds, both in ON phase and OFF phase, are almost linear with the number of active servers. The threshold to turn on one server in OFF phase is smaller than that in ON phase, indicating that the optimal action in OFF phase with no active server is to turn on service sooner to reduce the delay cost. The gap between other thresholds in ON phase and OFF phase, which correspond to turning on more than one servers, is constant. Moreover, servers are turned on more aggressively in ON phase with at least one active servers. The turn-off thresholds are shown in Fig. \ref{fig_off}. It is observed that the optimal action is not to turn off all the servers until the queue is emptied in OFF phase. Compared with the turn-on thresholds, the optimal action is to turn off servers only when the queue length is relatively quite small, and the servers are turned off very quickly when the queue length decreases beyond a certain point (about $10$ jobs in Fig. \ref{fig_off}). 
    \section{Conclusions}
    In this letter, we prove that the optimal sleeping mechanism with MMPP arrival and multiple servers is queue-threshold-based. This result settles a conjecture in the literature and extends to MMPP and multiple-server scenario. Through numerical results, it is shown that the optimal sleeping mechanism with multiple servers exhibits a slow activation, rapid and late (only when the queue length is quite small) shutdown feature. 
    
    \appendix
	\begin{lemma}[Partial submodular condition]
	\label{lm_1}
	If $\forall$ $S$, $W$, $t$, and $Q_1\leq Q_2$, $a_1\geq a_2$, 
	\begin{IEEEeqnarray}{rCl}
	    && u_t(S,Q_2,W,a_1)-u_t(S,Q_1,W,a_1) \leq u_t(S,Q_2,W,a_2)-u_t(S,Q_1,W,a_2),
	\end{IEEEeqnarray}
	the optimal policy is a monotone policy.
	\end{lemma}
	\begin{IEEEproof}
	Given $\forall$ $S$, $W$, and $t$, define 
	\begin{equation}
	g(Q)\triangleq \argmin_{a\in\{0,...,M\}}u_t(S,Q,W,a).    
	\end{equation} 
	Then $\forall$ $Q_1\leq Q_2$, 
	\begin{IEEEeqnarray}{rCl}
	&&u_t(S,Q_1,W,\min[g(Q_1),g(Q_2)])-u_t(S,Q_1,W,g(Q_1)) \nonumber\\
	& = &u_t(S,Q_1,W,g(Q_2))-u_t(S,Q_1,W,\max[g(Q_1),g(Q_2)])\nonumber\\
	&\leq &u_t(S,Q_2,W,g(Q_2))-u_t(S,Q_2,W,\max[g(Q_1),g(Q_2)])\leq 0.
	\end{IEEEeqnarray}
	This implies that $\min[g(Q_1),g(Q_2)]=g(Q_1)$, and thus $g(Q_1)\leq g(Q_2)$.
	\end{IEEEproof}
	\bibliographystyle{ieeetr}
	\bibliography{ibp}
\end{document}